\documentclass[prd,nofootinbib]{revtex4}
\def\journal#1, #2, #3#4, #5#6#7#8    {
    {#1~} {#2}  (#5#6#7#8) #3#4}

\def\prd{\journal Phys. Rev. D, }

\def\rmp{\journal Rev. Mod. Phys., }

\def\npb{\journal Nucl. Phys. B, }
\def\plb{\journal Phys. Lett. B, }

\def\ijmpa{\journal Int. Jour. Mod. Phys. A, }

\def\jmp{\journal J. Math. Phys., }
\def\jhep{\journal J. High Energy Phys., }

\newcommand{\beq}[1]{\begin{equation}\label{#1}}
\newcommand\eeq{\end{equation}}
\newcommand{\ba}[1]{\begin{eqnarray}\label{#1}}
\newcommand{\baa}{\begin{eqnarray}}
\newcommand\ea{\end{eqnarray}}
\newcommand{\bee}{\begin{equation}}

\def\nn{\nonumber \\}

\def\o{\omega}

\newcommand{\h}{Hamiltonian}

\newcommand{\B}[1]{{\bf #1}}
\def\hlf{\frac{1}{2}}

\newcommand{\nc}{noncommutative}

\begin{document}

\title{Harmonic oscillator on noncommutative spaces}

\author{Ivan Dadi\'c}
\email{dadic@thphys.irb.hr}
\author{Larisa Jonke}
\email{larisa@irb.hr}
\author{Stjepan Meljanac}
\email{meljanac@irb.hr}
\affiliation{Theoretical Physics Division,\\
Rudjer Bo\v skovi\'c Institute, P.O. Box 180,\\
HR-10002 Zagreb, Croatia}


\begin{abstract}
A generalized harmonic oscillator on noncommutative spaces is considered.
Dynamical symmetries  and physical 
equivalence of \nc \ systems with the same energy spectrum are investigated 
and discussed.
General solutions of three-dimensional \nc \ harmonic oscillator are  found 
and classified according to dynamical symmetries. 
We have found conditions under which three-dimensional \nc \ harmonic 
oscillator can be represented by ordinary, isotropic harmonic oscillator 
in effective  magnetic field.
\end{abstract}
\vspace{1cm}

\maketitle

\section{Introduction}

Recently, it has been
realized that noncommutatative geometry plays a  distinguished role in the 
formulation of string theory \cite{witt,sw}  and M-theory \cite{sch}.
It has been shown that, in a certain limit, the  entire string
dynamics can be  described by minimally
coupled gauge theory  on noncommutative space \cite{sw}.
Noncommutative field theory \cite{rev} 
has been constructed by introducing the Moyal 
product in the space of ordinary functions, and by defining field theory in 
quantum phase space. Equivalence  between these two approaches has been 
clarified in Ref.\cite{als}.

Presumably,   \nc \ effects are important at very high 
energies. Nevertheless, we could observe high-energy effects in the low-energy 
effective action, or we could use noncommutative geometry in constructing 
the low-energy effective action.
For example, the noncommutative Chern-Simons gauge theory, 
represented as a matrix theory of
elementary charges, has been used to describe quantum Hall physics \cite{qhe}.
Phenomenological aspects of spacetime noncommutativity have been analyzed
using a \nc \ extension of the standard model \cite{w}.
Recently, it was proposed to use synchrotron radiation to search for 
experimental evidence of the effects of noncommutativity \cite{ior}.
It has  turned out that in many proposals to test the hypothetical 
spacetime noncommutativity, it is 
sufficient to consider only quantum mechanical approximation \cite{gam,ab}.

Subsequently,  quantum mechanics
on noncommutative 
spaces has been extensively studied
\cite{nair,nairp,jell,torus,bell,smyr,sma2,sma3,esp,mm,lm,gamboa,un,ph,rb,ad,luk}.
In Ref.\cite{nas} two of us 
have presented  a unified approach to representations of
\nc \ quantum mechanics in arbitrary dimensions and have given 
conditions for physical equivalence of 
different representations.
Also, we have shown that there exist two physically distinct phases
in arbitrary dimensions that are connected by a discrete duality transformation.
Furthermore,  we have discussed symmetries in  phase space and the 
dynamical symmetry of a physical system, and shown how these symmetries are 
affected by change in commutation relations.

Throughout the paper we  assume that the time coordinate is commutative,
since otherwise we would be forced to modify the usual scheme of 
quantum mechanics \cite{li}.
We analyze a noncommutative harmonic oscillator (ho) in detail, 
especially three-dimensional case. 
Various deformations of the 
harmonic oscillator have been discussed in the literature,  including the
ho in the quantum group framework \cite{fiore}, the ho with minimal 
length uncertainty relations \cite{Kempf}, the ho in \nc \ 
spaces \cite{nairp,smyr,sma2,sma3}, 
the (super)ho on $CP^N$ \cite{bels}, just to mention the latest.
The overcomplete symmetry of the ho enables one to exactly solve the oscillator 
problem even in some deformed cases. Especially, the ho is the only 
exactly solvable model in \nc \ quantum mechanics.
As for applications, the ho represents a prototype of a physical system in 
every branch of physics, but  the most exciting  applications 
are probably Bose-Einstein condensation and quantum Hall effect
in various dimensions.

Surprisingly, the simplest physically relevant system, the three-dimensional 
ho, have not yet been  analyzed in detail. There exist some results concerning
very special choice of \nc \ parameters \cite{sma3,smyr,esp}.
The most general parametrization of the  \nc \ three-dimensional ho is 
nontrivial extension of the two-dimensional case, what can be best 
seen from our analysis of the dynamical symmetry of this \nc \ system. 

The plan of the paper is the following. In Section II we will introduce the 
general formalism used in the paper. Then, in  Section III we will 
present some results concerning harmonic oscillator on \nc \ space in 
arbitrary dimensions. The main results of the paper are related 
to the case of three-dimensional harmonic oscillator and are 
presented in Sect. IV. There, we have presented conditions under which
three-dimensional \nc \ harmonic 
oscillator can be represented by ordinary, isotropic harmonic oscillator
in the effective  magnetic field, see Refs.\cite{sma2,nairp} 
for two-dimensional case. 
And finally, in the last section we summarize 
our results.

The conventions we use in this paper are the following: summation over 
repeated indices is assumed, small latin 
indices $i,j,...$ go over configuration space dimensions, capital latin 
indices $\scriptstyle{I},\scriptstyle{J},...$  go over phase space dimensions,
bold face fonts denote vectors, prime denotes physical quantity in phase II of 
parameter space.

\section{Noncommutative spaces in arbitrary dimensions}

Let us define D-dimensional \nc \ coordinate operators $X_1,\ldots,X_D$ and the
corresponding \nc \ momentum operators $P_1,\ldots, P_D$ with commutation 
relations 
\baa\label{cr}
&& [X_i,X_j]=i\theta_{ij}, \;\;\;[P_i,P_j]=i B_{ij},\;i,j=1,\ldots,D\nn
&& [X_i,P_i]=ih_i,\;\;\;[X_i,P_j]=i\left\{\begin{array}{cc}
                                                \phi_{ij}& i<j \\
					-\psi_{ij}& i>j \end{array}\right.
\ea						
and   $X_i,\;P_i$ are hermitean operators,
$ X_i^{\dagger}=X_i, \;P_i^{\dagger}=P_i .$
The $2D-$ dimensional \nc \ phase  space is described by variables 
$(U_1,U_2,\ldots,U_{2D})=(X_1,P_1,\ldots,X_{D},P_{D})$ satisfying
\baa\label{mcr}
&& [U_I,U_J]=i M_{IJ},\;\;\; U_I^{\dagger}=U_I,\;{\scriptstyle I,J}=1,\ldots,2D,\nn
&& M_{JI}=-M_{IJ},\;\;\; (M_{IJ})^{\dagger}=M_{IJ}.\ea
Generally, $M_{IJ}$ is an operator depending on $U_I$. However  the Jacobi 
identities $[U_I,[U_J,U_K]]+{\rm cycl}.=0$ restrict  the  choice of 
the operators $M_{IJ}$.
An important physical requirement is that in the 
limit $\theta_{ij},B_{ij},\phi_{ij},
\psi_{ij} \rightarrow 0$ and $h_i\rightarrow\hbar$, we should obtain canonical 
variables $u_I$ 
of ordinary quantum
mechanics in a smooth manner.

If we start with an arbitrary real, nonlinear, and regular 
(invertible) mapping $U_I=U_I(u_J)$ and $u_J=u_J(U_I)$ connecting 
noncommuting and commuting phase-space variables, we obtain 
$[U_I,U_J]=i f_{IJ}(u_K)=i M_{IJ}(U_L)$ satisfying all the above restrictions
including the Jacobi identities.
The 
reverse is not true globally. Namely, by postulating the matrix $M$ satisfying
the Jacobi identities
it is not clear 
whether the above mentioned mapping exists. 
However, one can start from Eq.(\ref{mcr}) and 
find a {\it local} Darboux 
transformation $U_I=U_I(u_J)$, and $u_J=u_J(U_I)$.

The simplest examples of \nc \ spaces are i) $[X_i,X_j]=i\theta_{ij}$, with
$\theta_{ij}$ c-numbers, ii) $[X_i,X_j]=ic_{ijk}X_k$, Lie-algebra type,
iii) $ [X_i,P_j]=i[\delta_{ij}(1+\beta \B P^2)+\beta'P_iP_j]$, introducing 
minimal length uncertainty relations \cite{Kempf},
iv) $X_iX_j=R_{ij,kl}X_kX_l$, for example, the Manin plane, 
etc.
Interesting and important physical questions are:
i) What are the symmetries of such spaces and the corresponding conservation 
laws?,
ii) How can one define  classical and quantum physics 
on such spaces in a consistent way?,
iii) What are physical consequences of noncommutativity,
i.e.,  what are deviations from  physics on 
ordinary spaces?
iv) What are physical applications and the role of singular spaces
(example - \nc \ Landau problem)?


In the rest of the paper we  concentrate on a simple \nc \ space 
with c-number commutators $M_{IJ}$, Eq.(\ref{mcr}). Note that $\det M\geq 0$. 
Singular spaces (spaces at the critical point in parameter space) are 
defined by $\det M=0$, and regular (nondegenerate) spaces have $\det M$ 
strictly positive. The antisymmetric, real matrix $M$ is defined by $D(2D-1)$
parameters. It can be brought to a universal, block-diagonal  form 
\baa\label{uni}
 R^T M R =\left(\begin{array}{ccccc}
           0& \o_1& &  & \\
	  -\o_1&0& & & \\
            & &0&\o_2 & \\
            & &-\o_2&0& \\
	     & & & & \ddots \end{array}\right),\ea
where R is an orthogonal matrix with $\det R=1$, and $\pm\o$'s are real numbers, 
eigenvalues of the matrix $(iM)$. The characteristic equation is of order $2D$ and
contains only even powers in $\o$:
$$\prod_{i=1}^D (\o^2-\o_i^2)=0.$$
The product of eigenvalues 
$\kappa=\prod\o_i >0$ defines phase I, smoothly  connected with
ordinary space, and $\kappa<0$ defines phase II in parameter space.
In phase I we choose all eigenvalues positive, and in phase II we choose
$\o_D<0$ and  eigenvalues $\o_1,\o_2,\ldots,\o_{D-1}$ positive. 
The connection between 
the two phases is established by flip $F$ in the phase-space variables 
$X_D\leftrightarrow P_D$, i.e., $\o_D\leftrightarrow -\o_D$:
\baa\label{flip}
F=\left(\begin{array}{ccccccc}
1&0& & & & &\\
0&1& & & & &\\
 & &\ddots& & & &\\
 & & &1&0 & & \\
 & & &0&1& & \\
 & &  & & &0&1\\
  & && & &1&0\end{array}\right).\ea
We define the duality relations between 
the two phases, $|\o_i|=|\o_i'|, \;\kappa 
=-\kappa'$ \cite{nas}. These relations connect the physical systems with
$M'=R'FR^TMRFR'^T$, with $R,R'\in SO(2D)$.
A singular space is characterized by the degree of the eigenvalue $\o=0$.

In the rest of the paper we assume $\o_i\neq 0$.
The transformation $R$, Eq.(\ref{uni}), defines new variables $U_I^0=
R^T_{IJ}
U_J$, i.e., $(X_i^0,P_i^0)$ with the commutator
$$[X_i^0,P_j^0]=i\o_i\delta_{ij},$$
and all other commutators being zero. 
We can further transform the variables to ordinary canonical ones 
$u_I=(D^{-1})_{IJ}U^0_J$, or $u_I=(FD^{-1})_{IJ}U^0_J$ in phase 
II,
where the matrix D is $D={\rm diag}(\sqrt{\o_1},\sqrt{\o_1},
\ldots,\sqrt{\o_{D-1}},\sqrt{\o_{D-1}},\sqrt{|\o_D|},\sqrt{|\o_D|})$:
\baa\label{D}
&& x_i=X_i^0/\sqrt{\o_i},\;\;p_i=P_i^0/\sqrt{\o_i}, \forall i,\;\;\; 
{\rm in \;\;phase\;\; I}\nn
&& \left.\begin{array}{l}
x_i=X_i^0/\sqrt{\o_i},\;\;p_i=P_i^0/\sqrt{\o_i}, i=1,\ldots,D-1,\\
x_D=P_D^0/\sqrt{|\o_D|},\;\;p_D=X_D^0/\sqrt{|\o_D|}\end{array}\right\}
 {\rm in\;\; phase\;\; II} .\ea
The transformation $RD$ ($RDF$) connecting the 
initial noncommuting coordinates $U_I$ with 
the canonical $u_I$ is invertible but not unitary.

Note that the matrix $M$, Eq.(\ref{mcr}), is 
invariant under a group of transformations isomorphic to $Sp(2D)$. 
Furthermore, the orthogonal matrix $R$ is unique up to the orthogonal 
transformations
preserving $M$ $(S^TMS=M)$.

In $D$-dimensions, angular momentum operators are generators of coordinate
space rotations, preserving $\B X^2=\sum_i X_i^2,\;\B P^2=\sum_i P_i^2,\;
\B X\B P=X_iP_i$:
\baa\label{dj}
\left[J_{ij},X_k\right]&=&i(\delta_{ik}\delta_{jl}-\delta_{il}\delta_{jk})X_l,\nn
\left[J_{ij},P_k\right]&=&i(\delta_{ik}\delta_{jl}-\delta_{il}\delta_{jk})P_l,
\;i,j,k,l=1,\ldots,D\nonumber,\ea
and generally,
$$[J_{ij},U_K]=i(E_{ij})_{KL}U_L,\;{\scriptstyle K,L}=1,\ldots 2D.$$
For a regular matrix $M$, we can construct the angular momentum generators
$J_{ij}=-\frac{1}{2}(E_{ij}M^{-1})_{KL}U_KU_L$
only if
$[E_{ij},M]=0$, for all $i,j=1,\ldots,D$.
In $D=2$ case the angular momentum $J$ can be constructed only if 
$h_1=h_2=h$ and $\psi=\phi$
\bee\label{opci}
J=\frac{1}{h^2-\theta B+\phi^2}\left\{ h(X_1P_2-X_2P_1)
+\frac{B}{2}(X_1^2+
X_2^2)+\frac{\theta}{2}(P_1^2
+P_2^2)-\phi(X_1P_1+X_2P_2)
\right\}.
\eeq
For $D\geq 3$ there is no $SO(D)$ symmetry and we cannot construct 
all $D(D-1)/2$ angular momentum generators $J_{ij}$. There is at most
$\left[\frac{D}{2}\right]$ generators of rotations in mutually commuting 
\nc \ planes.

\section{Isotropic \nc \ oscillator in arbitrary \mbox{dimensions}}

The system is defined by the \h \
\bee\label{hd}
H=\frac{1}{2}\sum_{i=1}^D(P_i^2+X_i^2)=\frac{1}{2}\sum_{I=1}^{2D} U_I^2,\eeq
where the constants $\hbar$, $m$, and $\o$ are absorbed in phase-space variables.
The \h \ (\ref{hd}) is invariant under  $O(2D)$ transformations. The commutation
relations are described by the  c-numbers $M_{IJ}$.
We represent this system in terms of the canonical variables $u_I$, Eq.(\ref{D}), 
\baa\label{hcan}
H=\frac{1}{2}\sum_{I=1}^{2D} U_I^{0\;2}=\left\{ \begin{array}{cc}
\displaystyle\frac{1}{2}\sum_{I=1}^{2D}(Du)_I^2& {\rm phase\;\; I}\\
\displaystyle\frac{1}{2}\sum_{I=1}^{2D}(DFu)_I^2& {\rm phase\;\; II}\end{array}
\right\}=\frac{1}{2}\sum_{i=1}^D |\o_i|(p_i^{2}+x_i^{2}).\ea 
The matrix $F$, Eq.(\ref{flip}), represents a
discrete transformation connecting the two phases.
The energy spectrum is 
\bee\label{ean}
E_{n_1\ldots n_D}=\sum_{i=1}^D|\o_i|\left(n_i+\frac{1}{2}\right), 
\;n_i\in N_0.\eeq
The  \h \ (\ref{hcan}) and the energy spectrum (\ref{ean}) are equal in both
phases, for all $M$ with the same eigenvalues $|\o_i|$.
They correspond to the anisotropic oscillator in 
canonical variables. The initial isotropic \nc\ oscillator is not 
unitarily equivalent to the anisotropic oscillator in canonical 
variables, but all corresponding physical quantities of \nc \ harmonic 
oscillator can be 
uniquely determined. Only representations connected by
transformations preserving the commutation relations are physically
equivalent.

The degenerate energy levels for the  \h \ (\ref{hd}) are described by a
set of orthogonal eigenstates that transform according to an
irreducible representation of the dynamical symmetry group.
The dynamical symmetry group $G(H,M)$ is a group of all transformations
preserving both, commutation relations $M$ and the \h \ $H$.
For the fixed \h , the dynamical symmetry depends on $M$; so, by changing
the parameters of the matrix $M$ we can change $G(H,M)$ from $G_{min}(H,M)$
to $G_{max}(H,M)$.
For the \nc \ harmonic oscillator, the minimal dynamical symmetry group
is $[U(1)]^D$ and the maximal symmetry group is $U(D)$. 
Hence, different choices of $M$ correspond to different dynamical
symmetry. This can be viewed as a new mechanism of symmetry breaking
with the origin in (phase)space structure.   The underlying theory that would 
determine $M$ does not exist yet.

The generators of dynamical symmetry are quadratic in phase space
variables, i.e., of the type ${\cal G}=C_{IJ}U_IU_J$, where the
real coefficients
$C_{IJ}$ can be chosen to be
symmetric, $C_{IJ}=C_{JI}$. The generators  can be
determined as null-eigenvectors of the matrix $A_{IJ,KL}$:
$$[H,U_IU_J]=\sum_{K,L}A_{IJ,KL}U_KU_L=\frac{1}{2}\sum_K[M_{KI}(U_JU_K+U_KU_J)
+M_{KJ}(U_IU_K+U_KU_I)].$$
For the two-dimensional harmonic oscillator, we have calculated
\cite{nas}
the generators of $U(2)$ and generic $U(1)\times U(1)$ symmetry. There are
additional non-symplectic symmetries if $\o_i/\o_j$ are rational number
\cite{mrn}.

\subsection{$U(D)$ dynamical symmetry}

As we have already mention, the maximal dynamical symmetry group for the \nc \ 
harmonic oscillator is $U(D)$, and in that case the matrix $M$ has all 
eigenvalues identical, $\o_i=\o$. The general conditions are:
\bee\label{cond}
M^TM=\o^2\cdot1\!{\rm I}_{2D\times 2D}, \;\;M^T=-M.\eeq

In two dimensions we have solved Eqs.(\ref{cond}), and obtained the following 
conditions on noncommutativity parameters resulting in 
$U(2)$ symmetry of \nc \ harmonic oscillator in two dimensions \cite{nas}:
\baa\label{d2}
&& h_1=h_2=h,\; \theta= -B,\; \phi=\psi,\;{\rm in\; phase \;I},\nn
&& h_1=-h_2,\; \theta= B,\; \phi=-\psi,\;{\rm in\; phase \;II}.\ea
The angular momentum $J$  exists in phase I 
$$J=\frac{1}{h^2+\theta^2+\phi^2}\left[h(X_1P_2-X_2P_1)
-\phi(X_1P_1+X_2P_2)-\frac{\theta}{2}(X_1^2+
X_2^2-P_1^2-P_2^2)\right],$$
but does not exist in phase II.

For $D=4,5,6$, 
we can write  a simple family  
of antisymmetric  matrices $M$ that 
provide maximal $U(D)$ dynamical symmetry,  with fixed eigenvalue $\o$.
If $h_i=h$ for $i=1,\ldots,D$, then
\baa\label{m4}
M_{4}=\left(\begin{array}{cccc}
          J_0&A&B&A^T\\
	  -A^T&J_0&A&B\\
	  -B&-A^T&J_0&A\\
	  -A&-B&-A^T&J_0  \end{array}\right),\;\;
M_{5}=\left(\begin{array}{ccccc}
          J_0&A&B&B&A^T\\
         -A^T&J_0&A&B&B\\
         -B&-A^T&J_0&A&B\\
	 -B&-B&-A^T&J_0&A\\
        -A&-B&-B&-A^T&J_0  \end{array}\right),\ea
and
\baa\label{m6}
M_6=\left(\begin{array}{cccccc}
         J_0&A&B&B&B&A^T\\
	 -A^T&J_0&A&B&B&B\\ 
	  -B&-A^T&J_0&A&B&B\\
	  -B&-B&-A^T&J_0&A&B\\
	  -B&-B&-B&-A^T&J_0&A\\
	   -A&-B&-B&-B&-A^T&J_0 \\
	 \end{array}\right),\ea
where 
\baa\label{vidis}
J_0=\left(\begin{array}{cc}
          0&h\\ -h&0\end{array}\right),\;
A=\left(\begin{array}{cc}
          a&b\\ c&-a\end{array}\right),\;
B=\left(\begin{array}{cc}
          d&e\\ e&-d\end{array}\right).\ea
The fixed
eigenvalue is
\bee\label{B2}
\o^2=h^2+2a^2+b^2+c^2+(D-3)(d^2+e^2).\eeq
The real parameters $a,b,c,d,e,h$ satisfy the following relations:
\bee\label{B1}
2ad+h(b-c)+e(b+c)+(D-4)(d^2+e^2)=0,\;{\rm for\;D=4,5,6},\eeq
and in addition
\bee\label{B15}
a^2+bc+h(b-c)+d^2+e^2=0,\;{\rm for\;D=5,6}.\eeq
The most general case for $D=3$ is discussed in the next section.
Eqs.(\ref{m4}) and (\ref{m6}) represent  the parametrization in phase I, 
since there exists a smooth limit 
to  ordinary 
quantum mechanics.
The dual solution in phase II is $FMF$, and 
is obtained using the flip transformation (\ref{flip}).
More generally, there are solutions of Eqs.(\ref{cond}) with $h_i$ mutually 
different. 

The
\nc \ ho with $U(D)$ dynamical symmetry can be represented by ordinary
isotropic ho. These systems are not physically equivalent although they have 
the same energy spectrum.  
We can caluculate physical quantities of interest for 
ordinary ho, and,  using the mapping $RD$ ($RDF$), determine (uniquely)
corresponding physical quantities for \nc \ ho.
For harmonic oscillator defined on noncommutative space with $D\geq 3$,
there is no angular momentum generators.

\section{Harmonic oscillator in three dimensions}

The three-dimensional harmonic oscillator 
is the simplest, physically relevant system.
The most general matrix $M$ is 
\baa\label{ma}
M=\left(\begin{array}{cccccc}
 0&h_1&\theta_3&\phi_3&-\theta_2&-\phi_2\\
 -h_1&0&\psi_3&B_3&-\psi_2&-B_2\\
 -\theta_3&-\psi_3&0&h_2&\theta_1&\phi_1\\
 -\phi_3&-B_3&-h_2&0&\psi_1&B_1\\
  \theta_2&\psi_2&-\theta_1&-\psi_1&0&h_3\\
   \phi_2&B_2&-\phi_1&-B_1&-h_3&0\end{array}\right),\ea
with $\theta_{ij}=\varepsilon_{ijk}\theta_k$, and similarly for other 
parameters, $B_{ij},\phi_{ij},\psi_{ij}$, defined in Eq.(\ref{cr}). 
The critical points are determined by 
\baa
\det M &=&\left[h_1 h_2 h_3 + \sum_ih_i(\phi_i\psi_i-\theta_iB_i)+
\phi_1\phi_3\psi_2-\phi_2\psi_1\psi_3\right.\\ &+&\left. B_1(\psi_3\theta_2-
\psi_2\theta_3)+B_2(\psi_1\theta_3-\phi_3\theta_1)+B_3
(\phi_2\theta_1-\phi_1\theta_2)
\right]^2=0.\ea
The eigenvalues of the matrix  $iM$ are obtained from 
\bee\label{ch}
\o^6-\alpha \o^4+\beta \o^2-\gamma = 0,\eeq
where 
\baa\label{fact}
&&\alpha=\o_1^2+\o_2^2+\o_3^2=-\frac{1}{2}{\rm Tr} M^2,\nn
&& \beta=\o_1^2\o_2^2+\o_2^2\o_3^2+\o_1^2\o_3^2=\frac{1}{8}({\rm Tr} M^2)^2-
\frac{1}{4}{\rm Tr} M^4,\nn
&&\gamma= \o_1^2\o_2^2\o_3^2=\det M.\ea
Phase I is defined by $\kappa=\o_1\o_2\o_3>0$ and all $\o$'s 
positive, and phase 
II is defined by $\kappa<0$ and $\o_{1,2}>0,\o_3<0$. 
The duality relations connecting  the two phases with the same energy spectrum 
are 
$\o_1=\o_1',\;\o_2=\o_2',\;\o_3=-\o_3'$.
Singular spaces 
are characterized by i) $\o_1>0$ and $\o_2=\o_3=0$ and 
ii) $\o_1\geq\o_2$ and $\o_3=0$.

The spectrum of the \nc \ oscillator is the same as for the 
anisotropic harmonic 
oscillator, see Eqs.(\ref{hd}) and (\ref{hcan}):
\bee\label{ean3}
E_{n_1 n_2 n_3}=\o_1\left(n_1+\frac{1}{2}\right)+\o_2\left(n_2+\frac{1}{2}
\right)+|\o_3|\left(n_3+\frac{1}{2}\right),
\;n_1,n_2,n_3\in N_0.\eeq
All physical quantities for the \nc \ oscillator can be uniquely calculated 
knowing the transformation $RD$, i.e., $RDF$.

The generic dynamical symmetry is $[U(1)]^3$ when all eigenvalues $|\o_i|$ are 
mutually different. When two eigenvalues are the same, up to a sign, 
$\o_1=\o_2\neq \o_3$ or $\o_1\neq \o_2=\pm\o_3$, the dynamical symmetry is 
$U(2)\times U(1)$. The sign of $\o_3$ determines the phase. In the 
special case $\o_1=\o_2=\pm\o_3$, we have the $U(3)$ symmetry group.

\subsection{$U(3)$ dynamical symmetry}

In the case with $U(3)$ dynamical symmetry, 
$M^TM=\o^2\cdot 1\!{\rm I}_{6\times 6}$ and the most general 
parametrization 
of commutation relations with $h_i=h$, leading to $U(3)$ symmetry is
\baa\label{u33d}
M=\left(\begin{array}{ccc}
J_0&A&A^TR\\
-A^T&J_0&\varepsilon AR\\
-R^TA&-\varepsilon R^TA^T&J_0\end{array}\right),\ea
where the matrices $J_0$ and $A$ are given by Eq.(\ref{vidis}), 
$\varepsilon^2=1$,  and
\baa\label{male}
R=\left(\begin{array}{cc}
\cos\alpha&-\sin\alpha\\ \sin\alpha&\cos\alpha\end{array}\right).\ea
The  real parameters $a,b,c,h$ satisfy the following condition:
$$h(b-c)+\varepsilon(a^2+bc)=0,$$
and the common eigenvalue is $\o^2=h^2+2a^2+b^2+c^2$.
The matrix M, Eq.(\ref{u33d}), belongs to phase I. 
In phase II, $U(3)$ dynamical 
symmetry  can be parametrized by the 
matrix $FMF$, where $F$ is the flip matrix, Eq.(\ref{flip}).
More generally, there are solutions of Eqs.(\ref{cond}) with $h_i$ mutually
different.
For example, we can have
\baa\label{mass}
M=\left(\begin{array}{cccccc}
 0&h&\theta&\phi&0&0\\
-h&0&\phi&-\theta&0&0\\
 -\theta&-\phi&0&h&0&0\\
 -\phi&\theta&-h&0&0&0\\
  0&0&0&0&0&h_3\\
   0&0&0&0&-h_3&0\end{array}\right),\ea
with $\o^2=h_3^2=h^2+\theta^2+\phi^2$, also leading to the $U(3)$ dynamical
symmetry.

We have seen, Eqs.(\ref{u33d}) and (\ref{mass}), 
that there is a class of \nc \ isotropic oscillators with
$U(3)$ dynamical symmetry that are physically different. All of them are
described by the same \h \ $H=\sum U_I^2/2 $
and possess the same energy spectrum. They are connected by orthogonal 
transformations $R$ ($RF$). However, 
they are not unitarily equivalent, unless $[M,R]=0$. The easiest way
to see the difference is to consider the possibility of  saturation of 
uncertainty relations \cite{un,nas},
since $\Delta U_I\Delta U_J\geq |M_{IJ}|/2$.
Especially interesting is the case $M^TM=1\!{\rm I}_{6\times 6}$, where 
\nc \ ho $H=\sum U_I^2/2 $ and ordinary isotropic ho  $H=\sum u_I^2/2 $ 
have the same form, the identical spectrum, but have different 
matrix elements of observables.
In the latter case we can construct angular momentum operators, whereas
in the former (\nc) case this is not possible.

A special class of $U(3)$  invariant systems has  been
proposed in Ref.\cite{esp}, in order to retain $U(3)$ symmetry of
ordinary isotropic ho.
They have started with  the harmonic oscillator in terms of canonical variables
$H=\sum u_I^2/2$ and then transformed the system by the nonunitary
transformation
$U=RDu$, thus
obtaining $H=\sum(D^{-1}R^{-1}U)_I^2/2$. This system is quadratic in
$U_J$, but is not diagonal, and possesses $U(3)$ symmetry by construction.
But this is just   one special case in the  large class of $U(3)$-symmetric 
\nc \ harmonic  
oscillators that we have described in detail.

\subsection{Simple extension of two-dimensional ho}

There is a simple parametrization of the matrix (\ref{ma}), i.e., 
an extension of two-dimensional ho. 
Imagine we have \nc \ plane commuting with the rest of the space. In that 
case the matrix $M$ is  block-diagonal. There is one $4\times 4$ block 
representing four-dimensional phase space $(X_1,P_1,X_2,P_2)$ of \nc \ plane, 
and one $2\times 2$
block of the remaining coordinate $(X_3,P_3)$. 
We choose $B_i,\theta_i,\phi_i,\psi_i=0$, 
for $i=1,2$. This reduces to the most general two-dimensional harmonic 
oscillator
\cite{nas}. Specially, 
eigenvalues of the matrix $iM$ are:
\baa\label{opho}
&&\o_{1,2}=\frac{1}{2}\sqrt{(\theta-B)^2+(\phi+\psi)^2+(h_1+h_2)^2}
\pm \frac{1}{2}\sqrt{(\theta+B)^2+(\phi-\psi)^2+(h_1-h_2)^2},\nn
&& \o_3=h_3,\;\;\kappa=(h_1h_2-\theta B+\phi\psi)h_3.\ea
The phases are determined by the sign of $\kappa=\o_1\o_2\o_3$. The duality 
relations between system with the same energy spectrum in the two phases were 
constructed in Ref.\cite{nas}.

We can always represent \nc\ ho in terms of anisotropic ho in commuting 
coordinates. But, in two dimensions  
we can also represent 
the two-dimensional anisotropic oscillator as
the  two-dimensional isotropic oscillator in the effective magnetic field,
using a symplectic transformation between canonical
variables 
\bee\label{effh}
H=\hlf\sum_{i=1,2}[(p_i-A_i)^2+\o_{\rm eff}^2 x_i^2]=
\hlf \sum_{i=1,2}[p_i^2+(\o_{\rm eff}^2+B_{\rm eff}^2) x_i^2]
-\hlf B_{\rm eff}(x_1p_2-x_2p_1), 
\eeq
where $\B A=(-x_2,x_1,0)B_{\rm eff}/2$.
In phase I, effective
frequency and effective magnetic field are, for general two-dimensional
case:
\bee\label{effI}
\o_{\rm eff}^2=\o_1\o_2,\;\;
B_{\rm eff}=\o_1-\o_2=\sqrt{(\theta+B)^2+(\phi-\psi)^2+(h_1-h_2)^2}.\eeq
In phase II the corresponding physical quantities are
\bee\label{effII}
\o_{\rm eff}^2=\o_1|\o_2|=,\;\;B_{\rm eff}=\o_1-|\o_2|
=\sqrt{(\theta'-B')^2+(\phi'+\psi')^2
+(h_1'+
h_2')^2}.\eeq
This physical interpretation is possible only in two dimensions, or if \nc \
plane decouples from the rest of the higher-dimensional space.
The condition for the three-dimensional isotropic oscillator in the 
effective magnetic field along the third axis, $B_{\rm eff}=\o_1-|\o_2|$,
is $\o_1|\o_2|=\o_3^2=\o_{\rm eff}^2$.

Note that all three types of dynamical symmetry, i.e., $U(3)$, 
$U(2)\times U(1)$, and $[U(1)]^3$ are possible.

\subsection{Special parametrization in terms of $\mathbf{\Theta}$
and $\B B$}

For the rest of this section 
we choose $h_i=1$ and
$\phi_{ij}=\psi_{ij}=0$, i.e., we impose $[X_i,P_j]=i\delta_{ij}$.
We organize 
remaining parameters in the general matrix (\ref{ma})
in two vectors, $\mathbf{\Theta}=(\theta_1,\theta_2,\theta_3)$ 
and $\B B=(B_1,B_2,B_3)$. The condition for a 
critical point is $\kappa= 1-\mathbf{\Theta} \B B=0$, and the
sign of $\kappa$ determines 
the phase.
The coefficients in the characteristic equation (\ref{ch}) are
\baa\label{fact2}
&&\alpha=3+\mathbf{\Theta}^2+\B B^2\geq 3,\nn
&&\beta=3+\mathbf{\Theta}^2\B B^2+(\mathbf{\Theta}-\B B)^2\geq 3,\nn
&&\gamma= (1-\mathbf{\Theta} \B B)^2\geq 0.\ea
The coefficients $\alpha,\beta,\gamma$, and therefore the eigenvalues $\o_i$,
are invariant under rotation in three-dimensional space,  $\theta_{i}'=
R_{ij}
\theta_{j},\;B_i'=R_{ij}B_j,\; R\in O(3)$.

At the critical point, $\gamma=0$, i.e.,  $\mathbf{\Theta} \B B=1$. 
The  
equations (\ref{fact2}) imply $\alpha^2-4\beta>5$. There is only one 
zero, $\o_3=0$. The two remaining eigenvalues are 
$$\o_{1,2}^2=\frac{\alpha\pm\sqrt{\alpha^2-4\beta}}{2}.$$
At the critical point, the dynamical symmetry is $[U(1)]^3$.

\subsection{$\mathbf{\Theta}$ and $\B B$ collinear}

Especially interesting is the case when $\mathbf{\Theta}$ and $\B B$ 
are  collinear
and $\phi_{ij}=\psi_{ij}=0$. Then $\kappa=\o_1\o_2\o_3=
1-\mathbf{\Theta}\B B$, 
and the eigenvalues $\o_i$ can be reduced to the two-dimensional 
problem in the plane orthogonal to $\mathbf{\Theta}$. 
Namely, if we choose $\mathbf{\Theta}=(0,0,\theta)$ 
and $\B B=(0,0,B)$, we find a 
solution of characteristic equation (\ref{ch}):
\bee\label{solpar}
\o_{1,2}=\sqrt{\left(\frac{\theta-B}{2}\right)^2+1}\pm\left|\frac{\theta+B}{2}
\right|,\;\;\;\o_3=1.\eeq 
The solution can belong to phases I,II or to the singular case, depending on
$\kappa=\o_1\o_2\o_3$. 

The matrix $R$, defined in Eq.(\ref{uni}), can be written in this 
parametrization in the following  
form:
\baa\label{rr}
R =\left(\begin{array}{cccccc}
    \cos{\varphi}& 0& \sin{\varphi} & 0&0&0 \\
    0&\sin{\varphi} &0&\cos{\varphi} &0&0\\
    0&\cos{\varphi}&0& -\sin{\varphi}&0&0\\ 
    -\sin{\varphi}&0&\cos{\varphi}&0&0&0\\
    0&0&0&0&1&0\\
    0&0&0&0&0&1\end{array}\right)\ea
where we choose $\varphi\in(0,\pi/2)$, $\theta\geq 0$, $\theta+B\geq 0$ and
\bee\label{cs}
\cos{\varphi}=\frac{1}{\sqrt{1+(B+\o_2)^2}}=\frac{\o_2+\theta}
	 {\sqrt{1+(\o_2+\theta)^2}}=\sqrt{\frac{\o_1-B}{\o_1+\o_2}}.\eeq

The anisotropic oscillator with energy spectrum Eq.(\ref{solpar}) 
can  be represented 
as isotropic oscillator in $(x_1,x_2)$ plane in magnetic field orthogonal 
to that plane, see Eqs.(\ref{effI}) and 
(\ref{effII}). 
In given parametrization,  $h_1=h_2=1$, Eq.(\ref{effI}) 
reduces to \cite{sma3,rb}
\baa\label{eff}
&& \o_{\rm eff}^2=1-\theta B,\;\;B_{\rm eff}=\theta+B,\;
{\rm in\; phase\;I},\nn
&& \o_{\rm eff}^2=\theta' B'-1,\;\;B_{\rm eff}=\sqrt{(\theta'-B')^2+4},\;
{\rm in\; phase\;II}.\ea
Note that in the case $B=0$ (or $\theta=0$) we have three-dimensional
isotropic ho in effective magnetic field $B_{\rm eff}=\theta$ 
(or $B_{\rm eff}=B$),
since $\o_{\rm eff}^2=1=\o_3^2$.

Some simple examples are the following:

a) if $\B B=0$, the result belongs to phase I,
$$\o_{1,2}=\sqrt{1+\frac{\theta^2}{4}}\pm\frac{\theta}{2},\;\o_3=1.$$
We can construct symplectic transformation connecting anisotropic
oscillator with $3D$-isotropic oscillator in magnetic field with 
effective
frequency and effective magnetic field 
$$\o_{\rm eff}^2=1,\;\;B_{\rm eff}=\theta.$$

b) the choice $\theta=B$ leads to 
$$\o_{1,2}=1\pm\theta,\;\o_3=1,$$ and  it is singular for $\theta=1$, 
it belongs to phase I for $\theta<1$, and to phase II for $\theta>1$.
The effective
frequency and effective magnetic field of isotropic oscillator in magnetic 
field are
$$\o_{\rm eff}^2=1-\theta^2,\;\;B_{\rm eff}=2\theta<2\;{\rm in\; phase\; I},$$
and 
$$\o_{\rm eff}^2=\theta^2-1,\;\;B_{\rm eff}=2\;{\rm in\; phase\; II},$$
Also, for $\theta=B=2$, the eigenvalues are $\o_1=3,\o_2=-1,\o_3=1$ 
and we have $U(2)\times U(1)$ dynamical symmetry in phase II.

c) choosing $\theta=-B$  give us $U(2)\times U(1)$ dynamical symmetry 
in phase I,
since 
$$\o_1=\o_2=\sqrt{1+\theta^2},\;\o_3=1.$$
In this case, effective magnetic field cancels, $B_{\rm eff}=0$, and we have 
isotropic ho in two dimensions with effective frequency $\o_{\rm eff}^2=
1+\theta^2$.
The generators of the dynamical symmetry group were found in 
Ref.\cite{nas}.

An interesting physical example is  the \nc \ Landau problem \cite{nairp,ph}. 
In two 
dimensions it can be represented as a
 \nc \ harmonic oscillator with $\o\rightarrow 0$ and also
as a \nc \ harmonic oscillator with $\tilde \o\neq 0$,
at the  critical point $\tilde \theta\tilde B=1$.
The connection between parameters is
$\tilde\o^2\tilde\theta+1/\tilde\theta=B$.

With parametrization chosen in this subsection, 
the angular momentum in the conventional sense
(see the definition (\ref{dj}))
can be defined only in the plane orthogonal to $\mathbf{\Theta}$:
\bee\label{angul}
J_{12}=J_z=\frac{1}{1-\theta B}\left[X_1P_2-X_2P_1+\frac{B}{2}(X_1^2+X_2^2)+
\frac{\theta}{2}(P_1^2+P_2^2)\right].\eeq

Analyzing the characteristic  equation we find the following statements 
concerning dynamical symmetries with
nontrivial
noncommutative parameters ($\theta^2+B^2>0$).

i) There is no $U(3)$ dynamical symmetry in this 
parametrization. Namely, from $\o_1=1$ it  follows that 
$\theta=B$ and $\theta=-B$, and this is possible only 
for $\theta=B=0$.

ii) $U(2)\times U(1)$ dynamical symmetry can be realized in three 
ways. \\
First, there is a case described under c) above. There we have
$\o_1=\o_2>1,\;\o_3=1$  and $\mathbf{\Theta} $ and $\B B$ are antiparallel;\\
Secondly, we can have the following case in phase I:
$$\o_1=1+\frac{\theta^2}{\theta+1}>1,\;\o_2=\o_3=1.$$
For $\theta>0$ and $\theta+B>0$, $$B=-\frac{\theta}{\theta+1},$$
i.e., $\B B$ is antiparallel to $\mathbf{\Theta}$.
The effective frequency of two-dimensional 
isotropic ho is $\o_{\rm eff}=\o_1$ and 
effective magnetic field is $B_{\rm eff}=\theta^2/(\theta+1)$;\\
Finally, in phase II, for 
$\theta>1$ and $\theta+B>0$, we have 
$$\o_1=\frac{\theta^3+1}{\theta^2-1}>1,\;\o_2=-1,\;\o_3=1.$$
$\B B$ and $\mathbf{\Theta}$ are parallel and $$B=\frac{\theta}{\theta-1}.$$
The effective frequency of two-dimensional 
isotropic ho is $\o_{\rm eff}=\o_1$ and
effective magnetic field is $B_{\rm eff}=\theta^2/(\theta-1)-2$. Note that 
effective magnetic field $B_{\rm eff}$ need not be zero in order to have 
$U(2)\times U(1)$ dynamical symmetry.
Also note that  in above examples $\o_{\rm eff}>1$, hence there is no 
representation in terms of three-dimensional isotropic ho in effective 
magnetic field.

iii) in all other cases the generic dynamical symmetry is $[U(1)]^3$.

\subsection{Arbitrary position of $\mathbf{\Theta}$ and $\B B$}

In the more general case, when $\mathbf{\Theta}$ 
and $\B B$ are not collinear, the 
characteristic equation (\ref{ch}) 
leads to solutions that  are not easy to analyze. 
However, it is remarkable that the 
above statements for the dynamical symmetry group 
hold even when we extend the analysis to the case when 
$\mathbf{\Theta}$ and $\B B$ are not collinear.

\underline{Proposition:}
Let us assume that the noncommuting coordinates and momenta satisfy 
\bee\label{sati}
[X_i,X_j]=i\varepsilon_{ijk}\theta_k,\;\;
[P_i,P_j]=i\varepsilon_{ijk} B_k,\;\;,[X_i,P_j]=i\delta_{ij},\eeq
where $\theta_i$ and $B_i$ are real c-numbers, $\theta^2+B^2>0$, 
and consider the 
three-dimensional 
\nc \ isotropic oscillator (\ref{hd}).
Then, $U(2)\times U(1)$ dynamical symmetry is possible if 
and only if $\B B$ and $\mathbf{\Theta}$ are collinear, in three special cases. 
In all other cases, the 
generic dynamical symmetry is $[U(1)]^3$.

\underline{Proof:}
Let us express the invariants $\theta^2+B^2$, $\theta^2 B^2$, and 
$\mathbf{\Theta} \B B$ using $\alpha, \beta$, Eq.(\ref{fact}), and $\kappa$:
\baa\label{expres}
\theta^2+B^2 &=& \alpha -3, \nn
\theta^2 B^2&=& \beta-\alpha+2-2\kappa,\nn
\mathbf{\Theta} \B B &=& 1-\kappa.\ea
$U(2)\times U(1)$ dynamical symmetry implies that two out of 
three eigenvalues are equal, up to a sign, say 
$\o_1^2=\o_2^2=\o^2\neq \o_3^2$. Inserting $\o_1^2=\o_2^2=\o^2$ into 
(\ref{fact}), and using the inequalities
\bee\label{ine}
\theta^2+B^2>0,\;\; (\mathbf{\Theta} \B B)^2\leq\theta^2 B^2\leq 
\left(\frac{\theta^2+B^2}{2}\right)^2
,\eeq
we find the following inequalities:\\
From $\theta^2+B^2>0$ it follows
$$2\o^2+\o_3^2>3.$$
From $(\mathbf{\Theta} \B B)^2\leq\theta^2 B^2$ it follows
$$(\o_3^2-1)(\o^2-1)^2\leq 0.$$
From $4\theta^2 B^2\leq(\theta^2+B^2)^2$ it follows
$$(\o_3-1)^2[(\o_3+1)^2-4\o^2]\geq 0.$$
If $\o_3=1$, then $\o^2>1$, and if $\o^2=1$, then $\o_3^2>1$.
If $\o_3\neq 1$ and $\o^2\neq 1$, the above inequalities lead to a 
contradiction. From the above analysis it follows
that all $\o_1^2=\o_2^2$ solutions are possible if and 
only if $(\mathbf{\Theta} \B B)^2=\theta^2 B^2$, i. e., when 
$\B B$ and $\mathbf{\Theta}$ are collinear. 
When $\B B$ and $\mathbf{\Theta}$ are collinear, 
there are  three possible realizations of $U(2)\times U(1)$ symmetry 
as has already been shown in  Subsection C. $U(3)$ symmetry is 
not possible for this parametrization with $\theta^2+B^2>0$.

In conclusion, with parametrization $[X_i,P_j]=i\delta_{ij}$, 
only if $\B B$ and $\mathbf{\Theta}$ are collinear, 
$\o_3=1$ and it is possible to 
represent \nc \ ho  as ordinary $2D$ isotropic oscillator in effective 
magnetic field. But, \nc \ ho can be represented as $3D$ ordinary isotropic
oscillator in effective magnetic field $B_{\rm eff}=\o_1-|\o_2|$
even when $\B B$ and $\mathbf{\Theta}$ 
are not 
collinear, provided $\o_1|\o_2|=\o_3^2$. This condition is sufficient even for 
the most general parametrization of commutation relations $M$, Eq.(\ref{ma}).

\subsection{Application - Nilsson model}

As we have already noticed, 
different choices of $M$ correspond to different dynamical
symmetries and this can be interpreted as a new mechanism of symmetry breaking
with the origin in phase space structure.  There are possible applications
to bound states in atomic, nuclear, and particle physics. 
Let us briefly discuss the nuclear shell model, more specifically, 
the Nilsson model. 
The basis of the shell model in a finite nucleus is the assumption of an 
independent particle motion  within a mean field, and therefore 
the nuclear \h \ can be written as a sum of 
single-particle \h s over all active nucleons \cite{nill}.
The central binding potential for a nucleon may be  approximated by 
the harmonic 
oscillator well and in this case, the  single particle \h \ is
described by isotropic three-dimensional ho.
Phenomenological improvement of  such a simple model 
is done by breaking the rotational
symmetry of harmonic well,  leading to 
anisotropic ho with frequencies fitted from experiment.

We suggest that starting from the simple \h \ (\ref{hd})   defined on 
noncommutative space with $[X_i,P_j]=i\delta_{ij}$, 
we can formulate an effective theory that would account 
for the spectrum of low-lying excitations of a  nucleus. 
The parameters of the  matrix $M$, Eq.(\ref{ma}),
can be interpreted as parameters describing the 
background, namely, the mean field acting on a single nucleon. 
For example, we have shown in subsection D that the spectrum 
$e=\hbar\omega_1(n_1+1/2)+\hbar\omega_2(n_2+1)$
can be obtained from (\ref{hd}) when $\mathbf{\Theta}$ 
and $\B B$ are collinear.
In this case,  noncommutativity parameters have 
a clear physical interpretation,
representing  the preffered direction, thus breaking 
rotational invariance down to axial invariance. 
We have seen that we can define angular momentum 
in the plane orthogonal to $\mathbf{\Theta}$, Eq.(\ref{angul}),
so for particles 
with spin there is an 
additional term in \h \ (\ref{hd}), $H_{\rm int}\sim
\B J\cdot \B s$. 
There is a correction of order   $\theta$ in comparison to
commutative case. 
For large deformations we use the general
parametrization of the matrix $M$, Eq.(\ref{ma}).
The limit $\mathbf{\Theta},\B B\rightarrow 0$ reproduces 
rotationally symmetric, isotropic oscillator.
This being the effective theory,  the parameters
$\mathbf{\Theta}, \B B$ should be also determined from experimental results.
The difference  between our proposal and standard Nilsson model 
should be in the transition amplitudes. Namely, we can fit the \nc \ 
parameters to have  the desired energy spectrum (same as in commutative 
effective thoery), but then the  
matrix elements of observables would be different.

\section{Discussion and outlook}

We have considered a \nc , $O(2D)$ symmetric oscillator with constant 
(c-number) commutation relations $M_{ij}$ in $2D$-dimensional
nonsingular ($\det M\neq 0$) phase-space. 
There exist two, physically distinct,  phases, 
defined by 
$\kappa=\prod \o_i\;{\scriptstyle{>\atop <}}\;0$. 
If the  matrix  $M$ is block-diagonal 
($\det M=\prod \det M_i$), then it is characterized by the set of 
$\kappa_i$, corresponding to the  subspaces.
A discrete duality transformation 
connects two systems with the same energy spectra  in two different phases.
We have presented a unified approach for analyzing a \nc \ ho in 
arbitrary dimensions and both phases simultaneously. General construction
of transformations from \nc \ variables to canonical Darboux variables 
is presented. Starting from \h \ $H(U)$ and commutation relations $M$ 
and applying linear transformations,
we obtain different physical systems with the same energy spectrum and the 
same dynamical symmetry (up to isomorphism).
Namely, the noncommutative ho is mapped to an ordinary, commutative anisotropic ho
with the same spectrum as a  noncommutative ho. Since the transformation
$RD$ $(RDF)$ is not unitary, these two systems are not 
physically equivalent, althoght all physical quantities can be uniquely 
determined. These systems differ in matrix elements of observables, 
uncertainty relations and other physical properties. Only systems with the same 
energy spectra and the same commutation relations are physically (unitarily)
equivalent.

The dynamical symmetry of the \nc \ isotropic oscillator is $\prod U(D_i),\;
\sum D_i=D$. We have presented a family of matrices $M$ leading to the
maximal $U(D)$ symmetry of the ho in D-dimensional \nc \ space, for $D<7$. The 
\nc \ ho with $U(D)$ dynamical symmetry can be represented by ordinary 
isotropic ho. 

We have presented 
a detailed analysis of the three-dimensional \nc \ ho. 
Our main result is the parametrization of the 
matrix $M$ for different dynamical   
symmetry groups, $U(3),U(2)\times U(1), [U(1)]^3$. 
Especially, the most general conditions for maximal symmetry $U(3)$ are 
presented. We have shown that for a special parametrization of 
commutation relations, $h_i=1,\;\phi_{ij}=\psi_{ij}=0$, there 
is no $U(3)$ symmetry. Futhermore, the $U(2)\times U(1)$ dynamical symmetry 
is possible if and only if $\mathbf{\Theta}$ and $\B B$ are collinear (only in  
three particular cases). For an arbitrary angle, different from zero and $\pi$,  between
the vectors $\mathbf{\Theta}$ and $\B B$, the dynamical symmetry is $[U(1)]^3$.

We have found generally that three-dimensional \nc \ harmonic
oscillator can be represented by ordinary, $2D$ isotropic harmonic oscillator
in effective  magnetic field only in \nc \ plane that commutes with the third
dimension, or by $3D$ isotropic ho in the effective magnetic field 
provided $\o_1|\o_2|=\o_3^2$. 
Angular momentum operators in \nc \ spaces can be defined only as 
generators of rotations in \nc \ planes which mutually commute.
The physical interpretation
of \nc \ effects in quantum mechanics 
in higher dimensions is not yet clear and requires 
further investigations.

Acknowledgment

We thank M. Milekovi\'c, A. Smailagi\'c and D. Svrtan for useful discussions.
This work was supported by the Ministry of Science and Technology of the
Republic of Croatia under contract No. 0098002 and contract No. 0098003.


\begin{thebibliography}{99}
\bibitem{witt}
E. Witten, \npb 460, 33, 1996 .
\bibitem{sw}
N. Seiberg and  E. Witten, \jhep 09, 032, 1999 .
\bibitem{sch}
A. Connes, M. R. Douglas, and A. Schwarz, \jhep 02, 003, 1998 .
\bibitem{rev}
M. R. Douglas and N. A. Nekrasov, \rmp 73, 977, 2001 , and references therein.
\bibitem{als}
L. Alvarez-Gaume and S. R. Wadia,  Phys. Lett. B501 (2001) 319.
\bibitem{qhe}
L. Susskind, "The Quantum Hall Fluid and Non-Commutative Chern Simons Theory",
hep-th/0101029; A. P. Polychronakos, \jhep 04, 011, 2001 ;
L. Jonke and S. Meljanac, \jhep 01, 008, 2002 ; Phys. Rev. B66 (2002) 205313.
\bibitem{w}
X. Calmet, B. Jurco, P. Schupp, J. Wess, and M. Wohlgenannt,
Eur. Phys. J. C 23 (2002) 363.
\bibitem{ior}
P. Castorina, A. Iorio, and D. Zappal\`a, "Noncommutative Synchrotron", 
hep-th/0212238.
\bibitem{gam} 
H.~Falomir, J.~Gamboa, M.~Loewe, F.~Mendez, and J.~C.~Rojas,
Phys.\ Rev.\ D  66 (2002) 045018.
\bibitem{ab}
M. Chaichian, A. Demichev, P. Presnajder, M.M. Sheikh-Jabbari, and A. Tureanu,
\plb 527, 149, 2002 .
\bibitem{nair}
V.~P.~Nair, Phys.\ Lett.\ B 505 (2001) 249.
\bibitem{nairp}
V. P. Nair and A. P. Polychronakos, \plb 505, 267, 2001 .
\bibitem{jell}
A. Jellal, J. Phys. A 34 (2001) 10159.
\bibitem{torus}
B.~Morariu and A.~P.~Polychronakos,
Nucl.\ Phys.\ B 610 (2001) 531.
\bibitem{bell}
S. Bellucci, A. Nersessian and C. Sochichiu, \plb 522, 345, 2001 ,
S. Bellucci and  A. Nersessian, \plb 542, 295, 2002 .
\bibitem{smyr}
A. Hatzinikitas and I. Smyrnakis, \jmp 43, 113, 2002 .
\bibitem{sma2}
A. Smailagic and E. Spallucci, \prd 65, 107701, 2002 .
\bibitem{sma3}
A. Smailagic and E. Spallucci,
J.Phys. A35 (2002) L363-L368.
\bibitem{mm}
B. Muthukumar and P. Mitra, Phys. Rev. D66 (2002) 027701.
\bibitem{esp}
O. Espinosa and P. Gaete, "Symmetry in noncommutative quantum mechanics",
hep-th/0206066.
\bibitem{lm}
L.~Mezincescu, "Star Operation in Quantum Mechanics", hep-th/0007046.
\bibitem{gamboa}
J. Gamboa, M. Loewe, F. Mendez, J. C. Rojas, \ijmpa 17, 2555, 2002 .
\bibitem{un}
K. Bolonek, P. Kosinski, \plb 547, 51, 2002 .
\bibitem{ph}
C. Duval and P. A. Horv\'athy, Phys. Lett. B479 (2000) 284; 
P. A. Horv\'athy, Ann. Phys. 299 (2002) 128.
\bibitem{rb}
R. Banerjee, Mod. Phys. Lett. A 17 (2002) 631.
\bibitem{ad}
A. Deriglazov, "Noncommutative quantum mechanics as a constrained 
system", hep-th/0112053; \plb 530, 235, 2002 .
\bibitem{luk}
J. Lukierski, P. C. Stichel, and W. J. Zakrzewski, Annals Phys. 220 
 (1997) 224.
\bibitem{nas}
L. Jonke and S. Meljanac, "Representations of noncommutative quantum mechanics 
and symmetries", hep-th/0210042.
\bibitem{li}
Miao Li, \jhep 05, 033, 2002 .
\bibitem{fiore}
G. Fiore,  Int. J. Mod. Phys. A8 (1993) 4679;
U. Carow-Watamura and S. Watamura, Int. J. Mod. Phys. A9 (1994) 3989.
\bibitem{Kempf}
A.~Kempf, G.~Mangano, and R.~B.~Mann,
Phys.\ Rev.\ D {\bf 52} (1995) 1108, 
A. Kempf, J.\ Phys.\ A {\bf 30}, 2093 (1997), 
L.~N.~Chang, D.~Minic, N.~Okamura, and T.~Takeuchi,
Phys.\ Rev.\ D {\bf 65} (2002) 125027,
I. Dadi\'c, L. Jonke, and
S. Meljanac, "Harmonic oscillator with minimal length uncertainty relations and
ladder operators", hep-th/0210264, to appear in Phys. Rev. D.
\bibitem{bels}
S. Bellucci and A. Nersessian, "(Super)Oscillator on CP(N) and Constant Magnetic
 Field", hep-th/0211070.
\bibitem{mrn}
J.~F.~Carinena, G.~Marmo and M.~F.~Ranada,
J.\ Phys.\ A {\bf 35} (2002) L679.
\bibitem{nill}
W. Greiner, J. A. Maruhn, {\it Nuclear Models}, Springer 1996.
\end{thebibliography}
\end{document}